\documentstyle[dina4,epsfig]{report}
\begin{document}
\parskip=4pt plus 1pt
\textheight=8.7in
\newcommand{\beq}{\begin{equation}}
\newcommand{\eeq}{\end{equation}}
\newcommand{\beqa}{\begin{eqnarray}}
\newcommand{\eeqa}{\end{eqnarray}}
\newcommand{\no}{\nonumber}
\newcommand{\grts}{\greaterthansquiggle}
\newcommand{\lets}{\lessthansquiggle}
\newcommand{\ul}{\underline}
\newcommand{\ol}{\overline}
\newcommand{\ra}{\rightarrow}
\newcommand{\Ra}{\Rightarrow}
\newcommand{\ve}{\varepsilon}
\newcommand{\vp}{\varphi}
\newcommand{\vt}{\vartheta}
\newcommand{\dg}{\dagger}
\newcommand{\wt}{\widetilde}
\newcommand{\wh}{\widehat}
\newcommand{\dfrac}{\displaystyle \frac}
\newcommand{\fsl}{\not\!}
\newcommand{\ben}{\begin{enumerate}}
\newcommand{\een}{\end{enumerate}}
\newcommand{\bfl}{\begin{flushleft}}
\newcommand{\efl}{\end{flushleft}}
\newcommand{\ba}{\begin{array}}
\newcommand{\ea}{\end{array}}
\newcommand{\btab}{\begin{tabular}}
\newcommand{\etab}{\end{tabular}}
\newcommand{\bit}{\begin{itemize}}
\newcommand{\eit}{\end{itemize}}

\newcommand{\be}{\begin{equation}}
\newcommand{\ee}{\end{equation}}
\newcommand{\bearr}{\begin{eqnarray}}
\newcommand{\eearr}{\end{eqnarray}}
\newcommand{\bea}{\begin{eqnarray}}
\newcommand{\eea}{\end{eqnarray}}

\newcommand{\per}{\;\;.}
\newcommand{\mtiny}[1]{{\mbox{#1}}}
\newcommand{\mtinyn}[1]{{\mbox{\scriptsize{#1}}}}
\newcommand{\MS}{\mtiny{MS}}
\newcommand{\GeV}{\mbox{GeV}}
\newcommand{\MeV}{\mbox{MeV}}
\newcommand{\keV}{\mbox{keV}}
\newcommand{\ren}{\mtiny{ren}}
\newcommand{\kin}{\mtiny{kin}}
\newcommand{\hint}{\mtiny{int}}
\newcommand{\tot}{\mtiny{tot}}
\newcommand{\CHPT}{\mtiny{CHPT}}
\newcommand{\QED}{\mtiny{QED}}
\newcommand{\syst}{\mbox{syst.}}
\newcommand{\stat}{\mbox{stat.}}
\newcommand{\wave}{\mbox{wave}}
\newcommand{\co}{\; \; ,}
\newcommand{\nn}{\nonumber \\}
\newcommand{\fff}{\bar{f}}
\newcommand{\ffg}{\bar{g}}
\renewcommand{\theequation}{\arabic{equation}}
\renewcommand{\thetable}{\arabic{table}}

\newcommand{\fs}{\; \; .}

\newcommand{\bdm}{\begin{displaymath}}
\newcommand{\edm}{\end{displaymath}}

\newcommand{\nl}{\nonumber \\}

\thispagestyle{empty}

\renewcommand{\theequation}{\arabic{equation}}
\renewcommand{\thetable}{\arabic{table}}

\newcommand{\pl}{PL}
\newcommand{\spi}{s_\pi}
\newcommand{\ql}{QL}
\newcommand{\qn}{QN}
\newcommand{\pn}{PN}
\newcommand{\qq}{Q^2}
\newcommand{\mee}{m_l^2}
\newcommand{\thp}{\theta_\pi}

\hspace*{11cm} hep-ph/9412392\\
\vspace{1.2cm}
\begin{center}

{\Large{\bf{ The $\pi\pi$ scattering amplitude\\
in chiral perturbation theory}}}
\refstepcounter{chapter}
\label{pipi}
\thispagestyle{empty}

\vspace{2cm}
{\bf{J. Gasser}}

\vspace{.5cm}

Inst. Theor. Physik, Universit\"at Bern,    Sidlerstrasse 5, CH--3012 Bern

\vspace{0.5cm}
December 1994

\end{center}

\vspace{1cm}

\renewcommand{\thesection}{\arabic{section}}
\renewcommand{\thefigure}{\arabic{figure}}

\setcounter{equation}{0}
\setcounter{table}{0}
\setcounter{figure}{0}
\setcounter{subsection}{0}

\begin{center}{\bf Abstract}\end{center}

We discuss $\pi\pi$ scattering
in the framework of chiral perturbation theory. In
particular, we recall the predictions \cite{gale} for the  threshold parameters
and for the phase shift difference
$\delta_0^0-\delta_1^1$.

\vspace{1cm}

 \section{Notation}

The scattering amplitude for $\pi\pi$ scattering,
\bdm
\pi^i(p_1)\pi^k(p_2)\rightarrow\pi^l(p_3)\pi^m(p_4)\co
\edm
 reads
\bea
\langle\pi^m(p_4)\pi^l(p_3)\mtiny{out}|\pi^i(p_1)\pi^k(p_2)\mtiny{in}\rangle
&=&
\langle\pi^m(p_4)\pi^l(p_3)\mtiny{in}|\pi^i(p_1)\pi^k(p_2)\mtiny{in}\rangle\nn
&+& i(2\pi)^4  \delta^{(4)}(P_f-P_i)T^{ik;lm}(s,t,u)\co\nonumber
\eea
where $T^{ik;lm}(s,t,u)$ is a Lorentz invariant function of the standard
Mandelstam variables
\bea
s&=&(p_1+p_2)^2=4(M_\pi^2+q^2)\co\nn
 t&=&(p_3-p_1)^2=-2q^2(1-\cos{\theta})\co\nn
 u&=&(p_4-p_1)^2=-2q^2(1+\cos{\theta})\per\nonumber
\eea
$q\;(\theta)$ is the center-of-mass momentum (center-of-mass scattering angle).
On account of isospin symmetry, the amplitude $T^{ik;lm}(s,t,u)$ may be
expressed in terms of a single amplitude $A(s,t,u)=A(s,u,t)$,
\bdm
T^{ik;lm}(s,t,u)=\delta^{ik}\delta^{lm}A(s,t,u)+
                   \delta^{il}\delta^{km}A(t,s,u)+
                   \delta^{im}\delta^{kl}A(u,t,s)\per
\edm
To compare the calculated amplitude with  data on $\pi\pi$
scattering \cite{pipidata}, one expands the combinations with definite
isospin in the $s$-channel
\bea
T^0(s,t)&=&3A(s,t,u) +A(t,u,s)+A(u,s,t)\nn
T^1(s,t)&=&A(t,u,s)-A(u,s,t)\nn
T^2(s,t)&=&A(t,u,s) +A(u,s,t)\nonumber
\eea
into partial waves,
\bea
T^I(s,t)&=&32\pi
\sum_{l=0}^{\infty}(2l+1)P_l(\cos{\theta})t_l^I (s)\per\nonumber
\eea
Unitarity implies that in the elastic region $4M_\pi^2<s<16M_\pi^2$ the partial
wave amplitudes $t_l^I$ are described by real phase shifts $\delta_l^I$,
\bdm
t_l^I(s)=\left(\frac{s}{s-4M_\pi^2}\right)^{1/2}\frac{1}{2i}
\{e^{2i\delta_l^I(s)}-1\}\per
\edm
The behaviour of the partial waves  near threshold is of the form
\bdm
\mbox{Re}\;t_l^I(s)=q^{2l}\{a_l^I +q^2 b_l^I +O(q^4)\}\per
\edm
The quantities $a_l^I$ are referred to as the $\pi\pi$ scattering lengths.

\section{Data}

At low energies, the difference $\delta_0^0-\delta_1^1$
 may be
extracted in a
theoretically clean manner from data on $K_{\mbox{\scriptsize{{e4}}}}$ decays
\cite{ke4phase}. In the high-statistics
 CERN-Saclay experiment \cite{ross}, it has been measured in five energy
bins -- the values are displayed in table III in Ref. \cite{ross}
 and plotted in  figure
\ref{fig1} below. For earlier determinations of
$\delta_0^0-\delta_1^1$ from data on $K_{e4}$ decays, see table IV
 in Ref. \cite{ross}. Furthermore,  the measurement of the
lifetime of pionic atoms allows one to determine the difference
$|a_0^0-a_0^2|$ \cite{lifetime}. A corresponding experiment has been proposed
 at CERN \cite{czapek}.

\section{Theory}
In Ref. \cite{gale}, the amplitude $A(s,t,u)$ has been evaluated in the
framework of chiral $SU(2)\times SU(2)$ to one-loop accuracy,
\bea\label{amplit}
A(s,t,u)&=&\frac{s-M_\pi^2}{F_\pi^2}+B(s,t,u)+C(s,t,u)+O(E^6)\co\nn
B(s,t,u)&=&(6F_\pi^4)^{-1}\{3(s^2-M_\pi^4)\bar{J}(s)\nn
&+&[t(t-u)-2M_\pi^2t+4M_\pi^2u-2M_\pi^4]\bar{J}(t)+(t\leftrightarrow u)\}\co\nn
C(s,t,u)&=&(96\pi^2F_\pi^4)^{-1}\left\{2(\bar{l}_1-\frac{4}{3})
(s-2M_\pi^2)^2\right.\nn
&+&\left.(\bar{l}_2-\frac{5}{6})[s^2+(t-u)^2]+12sM_\pi^2(\bar{l}_4-1)+3M_\pi^4
(5-4\bar{l}_4-\bar{l}_3)\right\}\co
\eea
where $F_\pi$  is the pion decay constant,
and $\bar{l}_1,\ldots,\bar{l}_4$ are four of the ten low-energy parameters
that parametrize the effective lagrangian at next-to-leading order \cite{gale}.
 The
constants $\bar{l}_{1,2}$ can e.g. be determined by measuring the $D$-wave
scattering lengths $a_2^0$ and $a_2^2$ \cite{gale},
\bea
\bar{l}_1&=&480\pi^3F_\pi^4(-a_2^0+4a_2^2) +49/40+O(M_\pi^2)\co\nn
\bar{l}_2&=&480\pi^3F_\pi^4(a_2^0-a_2^2)
+27/20 +O(M_\pi^2)\co\nonumber
\eea
whereas the constant $\bar{l}_4$ is related to the scalar radius of the pion
\cite{gale},
 \bea
\bar{l}_4&=&\frac{13}{12}+\frac{8\pi^2F_\pi^2}{3}\langle r^2\rangle_S^\pi
+ O(M_\pi^2)\per\nonumber
\eea
The $S$- and $P$-wave threshold parameters are
\bea\label{escl}
a_0^0&=&\frac{7M_\pi^2}{32\pi F_\pi^2}\left\{1+
\frac{M_\pi^2}{3}\langle r^2\rangle_S^\pi
+\frac{200\pi
F_\pi^2M_\pi^2}{7}(a_2^0+2a_2^2)-\frac{M_\pi^2}{672\pi^2F_\pi^2}
(15\bar{l}_3-353)+O(M_\pi^4)\right\}\co \nn
b_0^0&=&\frac{1}{4\pi F_\pi^2}\left\{1+
\frac{1}{3}M_\pi^2\langle r^2\rangle_S^\pi
+40\pi
F_\pi^2M_\pi^2(a_2^0+5a_2^2)+\frac{39M_\pi^2}{64\pi^2F_\pi^2}
+O(M_\pi^4)\right\}\co\nn
 a_1^1&=&\frac{1}{24\pi F_\pi^2}\left\{1+
\frac{1}{3}M_\pi^2\langle r^2\rangle_S^\pi
+80\pi
F_\pi^2M_\pi^2(a_2^0-\frac{5}{2}a_2^2)+\frac{19M_\pi^2}{576\pi^2F_\pi^2}
+O(M_\pi^4)\right\}\co\nn
b_1^1&=&\frac{7}{2160\pi^3F_\pi^4}+\frac{10}{3}(a_2^0-\frac{5}{2}a_2^2)
+O(M_\pi^2)\per
\eea
The numerical values obtained by evaluating
 these improved low energy theorems
are given in  table \ref{tabsl} (in units of $M_{\pi^+}$).
 \begin{table}[t]
\begin{center}
\caption{Threshold parameters that are relevant in $K_{e4}$
experiments, in units of $M_{\pi^+}$.
\label{tabsl}}
\vspace{1em}
\begin{tabular}{lllll} \hline
&Soft&Experiment&Improved&size of\\
&pions&&low energy&correction\\
&&&theorems&3:1\\ \hline

$a_0^0$&$0.16$&$0.26\pm0.05$&$0.20\pm0.005$&$1.28$\\
$b_0^0$&$0.18$&$0.25\pm0.03$&$0.25\pm0.02$&$1.37$\\
$a_1^1$&$0.030$&$0.038\pm0.002$&$0.038\pm0.003$
&$1.26$\\
$b_1^1$&       &               &$(5\pm3)\times10^{-3}$&\\ \hline
\end{tabular}
\end{center}
\end{table}
In column 2 we give the soft pion predictions of Weinberg \cite{wein66},
obtained from the terms proportional to $F_\pi^{-2}$ in Eq. (\ref{escl}).
  The third
column contains the results of an analysis of the data as reported by Petersen
in the compilation of coupling constants and low-energy parameters
\cite{nagels}. The entries in the fourth column correspond to the
representation (\ref{escl}). Here, we have used the experimental
 $D$-wave
scattering lengths and the scalar radius of the pions as an input, together
with the value for $\bar{l}_3$ determined in \cite{gale}\footnote{To be
 more specific, we
use   $\langle r^2\rangle_S^\pi =0.60\pm
0.05{\mbox{fm}}^2 \; \cite{scalarpi},
 \bar{l}_3=2.9\pm 2.4 \; \cite{gale},
 a_2^0=(17\pm3)  \cdot
10^{-4}M_\pi^{-4} \; \cite{nagels},
 a_2^2=(1.3\pm 3)\cdot 10^{-4}M_\pi^{-4} \cite{nagels}, M_\pi=139.57$ MeV
$\; \cite{pdg}, F_\pi=92.4$ MeV$ \; \cite{pdg}$.}.

\noindent
{\underline{Remark:}}
The errors quoted in column 4 are obtained by adding the uncertainties
in $\langle r^2\rangle_S^\pi,a_2^0,a_2^2$ and in $\bar{l}_3$ in
quadrature.
They measure the accuracy, to which the first order corrections can be
calculated, and do not include an estimate of the contributions due to higher
order terms. Work to determine those reliably is in progress \cite{twoloops}.
 Note also, that the $S$-wave scattering lengths
vanish in the
chiral limit and we therefore have to expect relatively large electromagnetic
corrections to these quantities. To illustrate: if we use the mass of the
neutral pion rather than $M_{\pi^+}$, the prediction for $a_0^0$ is lowered by
0.016 (at a fixed value of $\bar{l}_{1,2,3,4}$). {\underline{End of remark}}.

Turning now to the energy dependence of the phase shifts, we note
that these may
be worked out from the explicit expression for the scattering amplitude given
above by use of   \cite{gameiss}
\bea
\delta_l^I(s)=(1-4M_\pi^2/s)^{1/2}\mbox{Re}\;t_l^I(s)
+O(E^6)\per\nonumber
 \eea
In the following, we concentrate on the phase shift difference
\bdm
\Delta=\delta_0^0-\delta_1^1\co
\edm
and obtain
\bea\label{loop}
\Delta&=&\Delta^{(2)}+\Delta^{(4)} + O(E^6)\co\nn
\Delta^{(2)}&=&
\frac{\rho M_\pi^2}{96\pi F_\pi^2}(5x+1)\co\nn
\Delta^{(4)}&=&
\rho M_\pi^4\left\{\frac{h(x)}{55296\rho^4x^2\pi^3 F_\pi^4}+
\frac{(5x+1)\langle
r^2\rangle_S^\pi}{288\pi F_\pi^2} +\frac{5}{48}(x^2+8x+12)a_2^0\right.\nn
&&+\left.\frac{25}{48}(7x^2-28x+24)a_2^2
-\frac{5\bar{l}_3}{1024\pi^3F_\pi^4}\right\}\co
\eea
where
\bea
h(x)&=&\rho^2(689x^3-4630
x^2+11396x-15240)x\nn
&&6\rho(50x^4-460
x^3+1319x^2-1028x-112)h_1(x)\nn
&&+36(3x^2-36x+106)h_1^2(x)\co\nn
h_1(x)&=&\ln\left\{\frac{1-\rho}{1+\rho}\right\}\;\;,\;\;
\rho=(1-4/x)^{1/2}\;\;,\;\;
 x=s/M_\pi^2\per
\eea
The quantity $\Delta^{(2)}$ stems from the leading order term
$(s-M_\pi^2)/F_\pi^2$ in Eq. (\ref{amplit}).
Numerical results
are displayed in the figures. In Fig.
\ref{fig1}, we show the data from Ref. \cite{ross}, together with the full
one-loop result $\Delta=\Delta^{(2)}+\Delta^{(4)}$ (solid line) and the
leading order term $\Delta^{(2)}$ (dashed
line). In Fig. \ref{fig2}, the various contributions to the next-to-leading
order term $\Delta^{(4)}$ are resolved. Notice that the
contribution from the low-energy constant $\bar{l}_3$ is very small.

For a discussion of the $\pi\pi$ amplitude
in the framework of generalized chiral  perturbation theory, see
 Ref. \cite{knecht}.

\section{Improvements at DA$\Phi$NE}
According to  Baillargeon and Franzini \cite{ke4phase}, DA$\Phi$NE will allow
  one
to determine the phase shift difference $\delta_0^0-\delta_1^1$
 with considerably higher precision than available now \cite{ross}.
It will, therefore, be of considerable interest to confront the above
 predictions with these  data. In particular, we note that a value of
$a_0^0=0.26$ is not compatible with the chiral prediction $a_0^0=0.20$.

\vspace{1cm}

\noindent
{\bf{Acknowledgements}}

\noindent
I thank Heiri Leutwyler for informative discussions.

\clearpage

\begin{figure}[t]
 \centerline{\epsfig{file=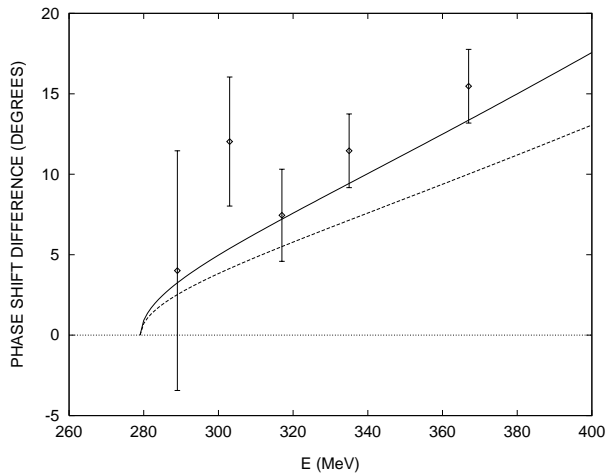,height=6cm}} \caption{
The phase shift difference $\Delta=\delta_0^0-\delta_1^1$ from the chiral
expansion. The data are from Ref. \protect{\cite{ross}}.
The solid line stands for the result  at one-loop accuracy, $\Delta=
\Delta^{(2)}+\Delta^{(4)}$, whereas the dashed line displays the leading order
term $\Delta^{(2)}$, see Eq. (\protect{\ref{loop}}) .\label{fig1}}
\end{figure}

\begin{figure}
\centerline{\epsfig{file=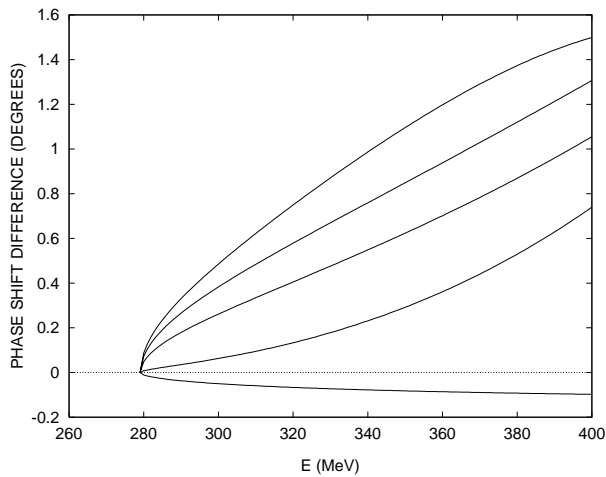,height=6cm}}
\caption{The various terms
in $\Delta^{(4)}$ according
 to Eq. (\protect{\ref{loop}}). From top to  bottom, the solid lines display
the contributions proportional to $h(x),\langle r^2\rangle_S^\pi, a_2^0,a_2^2$
and $\bar{l}_3$ in order. The sum of these terms generates the difference
between the solid and the dashed line in figure
\protect{\ref{fig1}}.\label{fig2}}
\end{figure}

\nopagebreak[4]

   \end{document}